\documentclass[10pt, conference]{IEEEtran}
\usepackage{ifthen}
\usepackage{amssymb}
\usepackage{color}

\newboolean{showcomments}
\setboolean{showcomments}{true}

\newsavebox\CBox

\ifthenelse{\boolean{showcomments}}
  {\newcommand{\nb}[2]{
    \fbox{\bfseries\sffamily\scriptsize\color{red}#1}
    {\sf\small$\blacktriangleright${\color{red}\textit{#2}}$\blacktriangleleft$}
   }
   
  }
  {\newcommand{\nb}[2]{}
   
  }

\newcommand{\ie}{\emph{i.e.},\xspace}

\newcommand{\eg}{\emph{e.g.},\xspace}

\newcommand{\etal}{\emph{et al.}\xspace}

\definecolor{gray50}{gray}{.5}
\definecolor{gray40}{gray}{.6}
\definecolor{gray30}{gray}{.7}
\definecolor{gray20}{gray}{.8}
\definecolor{gray10}{gray}{.9}
\definecolor{gray05}{gray}{.95}

\newlength\Linewidth
\def\findlength{\setlength\Linewidth\linewidth
  \addtolength\Linewidth{-4\fboxrule}
  \addtolength\Linewidth{-3\fboxsep}
}
\newenvironment{examplebox}{\par\begingroup
  \setlength{\fboxsep}{5pt}\findlength
  \setbox0=\vbox\bgroup\noindent
  \hsize=0.95\linewidth
  \begin{minipage}{0.95\linewidth}\normalsize}
  {\end{minipage}\egroup
  \textcolor{gray20}{\fboxsep1.5pt\fbox
    {\fboxsep5pt\colorbox{gray10}{\normalcolor\box0}}}
  \endgroup\par\noindent
  \normalcolor\ignorespacesafterend}

\usepackage{cite}
\usepackage{comment}
\usepackage{enumerate}
\usepackage{graphicx, color}                       
\usepackage[utf8]{inputenc}
\usepackage{paralist}               
\usepackage{tabularx,ragged2e}                            
\usepackage{balance}                          
\usepackage{multirow}     
\usepackage{multicol}            
\usepackage{fancyvrb}       
\usepackage{xspace}
\usepackage{amsmath}
\usepackage{hyperref}
\usepackage{cleveref}
\usepackage{lipsum}
\usepackage{tcolorbox}
\usepackage{booktabs}
\usepackage{makecell}
\usepackage{listings}
\usepackage{todonotes}
\usepackage{colortbl}

\usepackage{enumitem}
\usepackage[para]{footmisc}

\usepackage{listings}

\definecolor{mygreen}{rgb}{0,0.6,0}
\definecolor{mygray}{rgb}{0.5,0.5,0.5}
\definecolor{mymauve}{rgb}{0.58,0,0.82}
\definecolor{dkgreen}{rgb}{0,0.6,0}
\definecolor{gray}{rgb}{0.5,0.5,0.5}
\definecolor{mauve}{rgb}{0.58,0,0.82}
\definecolor{gray}{rgb}{0.4,0.4,0.4}
\definecolor{darkblue}{rgb}{0.0,0.0,0.6}
\definecolor{lightblue}{rgb}{0.0,0.0,0.9}
\definecolor{cyan}{rgb}{0.0,0.6,0.6}
\definecolor{darkred}{rgb}{0.6,0.0,0.0}
\definecolor{gray50}{gray}{.5}
\definecolor{gray40}{gray}{.6}
\definecolor{gray30}{gray}{.7}
\definecolor{gray20}{gray}{.8}
\definecolor{gray10}{gray}{.9}
\definecolor{gray05}{gray}{.95}

\lstdefinelanguage{XML}
{
  morestring=[s][\color{mauve}]{"}{"},
  morestring=[s][\color{black}]{>}{<},
  morecomment=[s]{<?}{?>},
  morecomment=[s][\color{dkgreen}]{<!--}{-->},
  stringstyle=\color{black},
  identifierstyle=\color{lightblue},
  keywordstyle=\color{red},
  morekeywords={plugin,groupId,artifactId,configuration,includes,include,excludes,exclude}
}

\newcounter{Finding}
\stepcounter{Finding}

\newcommand{\revised}[1]{\textcolor{black}{#1}}

\newcommand{\resquestion}[2]{ %
	\vspace{5pt} %
	\noindent\fcolorbox{black}{blue!05}{%
		\parbox{0.97\linewidth}{%
			\textbf{RQ$_{#1}$.} \emph{#2} %
		}%
	}%
	\vspace{5pt} %
}%

\begin{document}

\title{Machine Learning-Based Test Smell Detection}

\author{\IEEEauthorblockN{Valeria Pontillo\IEEEauthorrefmark{1}, Dario Amoroso d'Aragona\IEEEauthorrefmark{2}, Fabiano Pecorelli\IEEEauthorrefmark{2},}
\IEEEauthorblockN{Dario Di Nucci\IEEEauthorrefmark{1}, Filomena Ferrucci\IEEEauthorrefmark{1}, Fabio Palomba\IEEEauthorrefmark{1}}

\IEEEauthorblockA{vpontillo@unisa.it, dario.amorosodaragona@tuni.fi, fabiano.pecorelli@tuni.fi}
\IEEEauthorblockA{ddinucci@unisa.it, fferrucci@unisa.it, fpalomba@unisa.it}

\IEEEauthorblockA{
    \IEEEauthorrefmark{1}Software Engineering (SeSa) Lab --- University of Salerno, Fisciano, Italy
    }
    
\IEEEauthorblockA{
    \IEEEauthorrefmark{2}Tampere University --- Tampere, Finland
    }
    }
\maketitle
\thispagestyle{plain}
\pagestyle{plain}

\begin{abstract} 
	Context: Test smells are symptoms of sub-optimal design choices adopted when developing test cases. Previous studies have proved their harmfulness for test code maintainability and effectiveness. Therefore, researchers have been proposing automated, heuristic-based techniques to detect them. However, the performance of such detectors is still limited and dependent on thresholds to be tuned. 
	
	Objective: We propose the design and experimentation of a novel test smell detection approach based on machine learning to detect four test smells. 
	
	Method: We plan to develop the largest dataset of manually-validated test smells. This dataset will be leveraged to train six machine learners and assess their capabilities in within- and cross-project scenarios. Finally, we plan to compare our approach with state-of-the-art heuristic-based techniques.
	
	
\end{abstract}


\section{Introduction}
\label{sec:intro}
Test cases are the first barrier against software faults, particularly during regression testing~\cite{myers2011art}. 
Development teams rely on their outcome to decide whether it is worth merging a pull request~\cite{gousios2015work} or even deploying the system~\cite{beller2017oops}.
At the individual level, the developer's productivity is also partially dependent on the ability of tests to find real defects in production code~\cite{zhang2015assertions} and the timely diagnosis of the underlying causes~\cite{perez2017test}. Unfortunately, when developing test cases, programmers may apply sub-optimal implementation choices that could introduce test debt~\cite{kruchten2012technical}, namely potential design problems that lead to unforeseen testing and debugging costs for developers~\cite{maldonado2015detecting}.  
Test smells, \ie symptoms of poor design or implementation choices in test code~\cite{van2001refactoring}, represent one of the major sources of test debt~\cite{samarthyam2017understanding,tufano2016empirical}. Several empirical studies have focused on test smells to understand their properties~\cite{tufano2016empirical} and their impact on maintainability~\cite{spadini2018relation,bavota2015test,grano2019scented} and test effectiveness~\cite{grano2019lightweight}, by showing compelling evidence of the risks associated with the presence of test smells for software dependability.

For these reasons, researchers have investigated methods for automatically detecting test smells~\cite{garousi2018smells}.
Such techniques discriminate tests affected (or not) by a certain type of smell by applying detection rules that compare the values of relevant metrics extracted from test code against some empirically identified thresholds.
For instance, van Rompaey \etal~\cite{van2007detection} proposed a metric-based technique that computes several structural metrics (\eg number of production code calls made by a test case) and combines them into detection rules to highlight the likelihood of a test being smelly. A test is marked as smelly if the value overcomes a threshold.

Despite the effort spent by researchers so far, existing test smell detectors suffer from two key limitations.
First and foremost, they have limited detection capabilities, behaving similarly to a random guessing approach~\cite{van2007detection,greiler2013automated,palomba2018automatic}.
Secondly, their performance is strongly influenced by the thresholds used in the detection rules to discriminate between smelly and non-smelly tests~\cite{fernandes2016review,garousi2018smells}.
These restrictions threaten the practical applicability of these approaches. 

This registered report proposes the design of a novel test smell detector to overcome these limitations. The approach will be based on machine learning and employ structural and textual metrics to estimate the likelihood of a test being smelly.
Besides avoiding the need to combine metrics using detection rules, a machine learning approach also avoids the problem of selecting thresholds, thus representing a promising solution to alleviate the limitations of heuristic-based techniques.
Our approach is instantiated for the detection of four test smell types, \ie \emph{Eager Test}, \emph{Mystery Guest}, \emph{Resource Optimism}, and \emph{Test Redundancy}.
\revised{Afterward, we propose an empirical evaluation plan to assess the performance of the devised detector on a new dataset of \textsc{Java} projects---which we will manually build, publicly releasing the largest manually-crafted dataset of test smells to date \cite{garousi2018smells}---and compare its performance with three state-of-art heuristic-based techniques.}

\section{Related Work}
\label{sec:related}

Investigations on the design of test code were originally pointed out by Beck~\cite{beck2003test}.
Van Deursen \etal~\cite{van2001refactoring} and Maszaros~\cite{meszaros2007xunit} defined catalogs of test smells along with their refactoring actions.
More recently, Greiler \etal~\cite{greiler2013automated} devised \textsc{TestHound}, a heuristic-based approach to identify six test smell types that was evaluated through semi-structured interviews.
Palomba \etal~\cite{palomba2018automatic} devised \textsc{Taste}, a test smell detector which leverages textual metrics (\eg the conceptual cohesion of test methods \cite{marcus2005conceptual}) to complement previous techniques and identify three test smell types. \revised{The detection rules proposed by Palomba \etal~\cite{palomba2018automatic}, were later implemented in \textsc{Darts}~\cite{lambiase2020just}, an \textsc{Intellij} plugin that makes \textsc{Taste} usable through a user interface.}
Peruma \etal~\cite{peruma2020tsdetect} proposed \textsc{tsDetect}, a test smell detector  that identifies 19 test smell types, including \emph{Assertion Roulette}, \emph{Eager Test}, and \emph{Lazy Test}. 
Pecorelli \etal~\cite{pecorelli2020vitrum} proposed \textsc{VITRuM}, a \textsc{Java} plugin to provide developers with static and dynamic test-related metrics and identify seven test smell types. \revised{Similarly, Wang \etal~\cite{wang2021pynose} proposed \textsc{PyNose}, a \textsc{Python} plugin to detect 17 test smells.}
Koochakzadeh \etal ~\cite{koochakzadeh2010tester} proposed \textsc{TeReDetect}, a tool that uses rules and dynamic metrics to detect \emph{Test Redundancy}, \ie a test that could be removed without impacting the test suite.
De Bleser \etal~\cite{de2019socrates} proposed \textsc{SoCRATES}, a fully automated tool that combines syntactic and semantic data to identify six test smells in \textsc{Scala} software systems. 
Our paper is complementary to these researches since it introduces a new, orthogonal method based on machine learning to identify test smells that will not require tuning thresholds.
Furthermore, we plan to conduct a large-scale empirical study on a manually-validated dataset, making our investigation the largest ever done in test smell detection research.
Other related work concerns the empirical analyses of test smells. Tufano \etal~\cite{tufano2016empirical} investigated the lifecycle of test smells, while Bavota \etal~\cite{bavota2015test} showed that test smells are highly diffused in software projects and impact the understandability of test code. Similar results were later achieved when considering automatically generated test cases~\cite{grano2019scented} and in software systems developed using the combination of \textsc{Scala} and \textsc{ScalaTest}~\cite{de2019assessing}. Furthermore, Spadini \etal~\cite{spadini2018relation} showed that test smells impact the maintainability of both test and production code. Spadini \etal~\cite{spadini2019test} also discovered that test-driven code reviews might help developers discover design flaws in test code. All these studies serve as motivation for our paper.

\revised{Based on the empirical evidence provided in the past, test smells represent a relevant threat to software reliability that should be promptly detected. We aim to employ machine learning (ML) algorithms, which have been previously used for code smell detection---the interested reader may find a comprehensive literature analysis on code smells in~\cite{azeem2019machine}.} 
\revised{Although code and test smells share a similar high-level definition, they do not share the same characteristics. It is, therefore, worth analyzing the main differences we expect compared to the previous researches on code smell detection. According to the literature available, ML-based code smell detection comes with three significant limitations concerning (i) data imbalance, (ii) subjectivity of code smell data, and (iii) a set of predictors that poorly contribute to the accuracy of the detection~\cite{pecorelli2019comparing}.}

\revised{As for the data imbalance limitation, previous literature has shown that test smells are more diffused than code smells, \eg Bavota \etal \cite{bavota2015test} found \emph{Eager Test} instances to affect around 35\% of test classes. Conversely, code smells typically affect a meager percentage of classes (\ie around 2\%) \cite{palomba2016diffusion}. Therefore, we think the limitation of data imbalance could have a lower significance when dealing with test smells. Nevertheless, we planned to investigate the use of data balancing to understand whether this additional step could benefit the models.}

\revised{Concerning subjectivity, we envision a strong relationship between test and code smells. The dataset we will build may suffer from the subjectivity of the authors who will make the validation; therefore, we will also employ external developers, as explained in \Cref{subsec:validation}.} 

\revised{As for the predictors, we will rely on metrics adopted by existing heuristic techniques. While we applied the same strategy as previously done for code smell detection \cite{pecorelli2019comparing}, heuristic techniques for test smell detection are typically more accurate than those dealing with code smell detection~\cite{aljedaani2021test}. Therefore, we expect the performance of our models not to be strongly influenced by this limitation.}

\section{Goals and Research Questions}
\label{sec:questions}

The \emph{goal} of the study is to evaluate the extent to which machine learning is suitable for test smell detection, with the \emph{purpose} of improving test code quality by removing detrimental design flaws.
The \emph{perspective} is of researchers and practitioners interested in understanding the performance and limitations of machine learning techniques for test smell detection.
Specifically, our paper is structured around four research questions (\textbf{RQ}s). 

\resquestion{1}{\revised{Which are the features that provide more information gain to a machine learning-based test smell detector?}}

\resquestion{2}{What is the performance of a machine learning-based test smell detector?}

\resquestion{3}{How does a machine learning-based test smell detector perform compared to heuristic-based approaches?}

\revised{With the first research question (\textbf{RQ$_1$}), we seek to understand which metrics contribute the most to the detection of test smells. These observations will be used to (i) quantify the predictive power of metrics and (ii) identify the most promising features to include in our machine learning approach. In \textbf{RQ$_2$} we run our machine learning approach against a manually-validated oracle of test smells (described later in this section) to quantify its detection performance capabilities. Afterward, with \textbf{RQ$_3$} we aim to compare the performance of our technique with the one achieved by state-of-the-art approaches based on heuristics: such validation will allow us to understand the actual value of a machine learning approach, \ie should it work worse than heuristic approaches, its usefulness would be limited, as practitioners might still found heuristic approaches more beneficial. Last but not least, we plan for an additional \emph{in-vivo} evaluation of the capabilities of the machine learning model. Should the results coming from \textbf{RQ$_2$} and \textbf{RQ$_3$} be sufficiently promising, we aim to investigate the extent to which the predictions made are useful for developers to diagnose and/or refactor test smells. Hence, the last research question will be:}

\resquestion{4}{\revised{What is the practitioners' perception of the test smells output by a machine learning-based test smell detector?}}

\revised{To design and report our empirical study, we will follow the empirical software engineering guidelines by Wohlin et al. \cite{wohlin2012experimentation}, other than the \emph{ACM/SIGSOFT Empirical Standards}.\footnote{Available at \url{https://github.com/acmsigsoft/EmpiricalStandards}.}}

\section{Dataset Construction}
\label{sec:dataset}

The first step of our investigation will be the creation of a manually-validated dataset of test smells. 

\subsection{Projects Selection}
\label{subsec:sel_proj}
\revised{We will first collect test data from a dataset of 70 open-source \textsc{Java} projects, publicly available on \textsc{GitHub}, and 51,549 test cases. These projects are part of a larger, popular dataset known as the International Dataset of Flaky Tests (IDoFT).\footnote{\url{https://mir.cs.illinois.edu/flakytests/}} The selection is driven by two main factors. First, we consider the entire set of test cases contained in these projects, \ie not only those labeled as flaky, to complement  IDoFT with additional information related to test smells. In this way, researchers will be provided with a unique database containing various test code-related issues, which would be beneficial to stimulate further research on test code quality. These projects are highly diverse: they have different characteristics, scopes, and sizes. For the lack of space, more detailed statistics on those projects are available in our online appendix~\cite{appendix}}.
\revised{Secondly, the rationale for using this dataset comes from previous observations made by Pontillo \etal~\cite{pontillo2021feasibility}. In their study, the authors ran a state-of-the-art test smell detector named \textsc{VITRuM}~\cite{pecorelli2020vitrum} and identified a high number of test smells, \ie they found that around 80\% of test cases are smelly. While we will not use automated tools to collect test smell data, the high diffuseness of test smells in the dataset suggests that it may be worth manually analyzing them.}

\subsection{Selecting Test Smells}
\label{subsec:smell_selection}
We have already performed a comprehensive literature analysis to extract all the test smells automatically detectable by the current techniques.
We started from the list of test smell detection tools reported in a systematic mapping study by Aljedaani \etal~\cite{aljedaani2021test}.
This study reports all the test smell detection tools available in the literature and the test smells they detect.
From an initial set of 22 tools, we included only those (i) supporting \textsc{Java} as a programming language, as the vast majority of tools use only \textsc{Java} as the target language, and (ii) relying on a metric-based approach, since machine learning classifiers require a set of metrics to be used as predictors.
This filtering phase led us to a final number of ten tools. 

Afterward, we analyzed each tool and extracted information about the test smells they detect and the metrics they use for the detection. As a machine learning-based classification would be meaningless if based on a single metric, we decided to include only test smells for which at least two metrics have been defined (more details about the metrics are reported in \Cref{sec:independent}). This led us to the selection of a set of six test smell types, namely \emph{Empty Test}, \emph{Eager Test}, \emph{Mystery Guest}, \emph{Sensitive Equality}, \emph{Resource Optimism}, and \emph{Test Redundancy}.

However, we noticed that detecting two of these smells was very trivial (\ie \emph{Empty Test} and \emph{Sensitive Equality}); therefore, the use of a machine learning-based approach would not lead to any detection performance improvement. \emph{Empty Test} is defined as \emph{``A test method that is empty or does not have executable statements''}; thus, a heuristic approach could objectively identify test cases that suffer from this issue. The same consideration also applies to \emph{Sensitive Equality}, which occurs when \emph{``an assertion has an equality check by using the toString method''}.
Based on the above consideration, we decided to discard these two test smells, resulting in a final set of four test smells reported in \Cref{tab:test-metrics} together with their definition.
\revised{Another discussion point concerns the \emph{Resource Optimism} smell. Given its definition, it is likely that information-flow or dynamic analyses might be potentially more suitable for detecting it. In this sense, a machine learning solution might be sub-optimal, yet we aim at experimenting with the extent to which it may provide valuable insights to detect the smell. These observations might be used to understand how the performance of machine learning compares to existing approaches and, perhaps, be later used by researchers to combine it with novel, more precise information flows or dynamic sources of information.}

\subsection{Collecting Test Smell Data}
\label{subsec:validation}
Considering the unavailability of a reliable dataset containing information about test smell occurrences, we plan to manually identify them in the dataset described in \Cref{subsec:sel_proj}.
The first two authors of the paper (from now on ``the inspectors'') will conduct the entire process to mitigate the subjectivity of the validation.
Given the impracticability of manually analyzing all 51,549 test cases, the process will be conducted on a statistically significant stratified sample of 12,550 test cases (confidence level = 99\%, margin of error = 1\%). The sample will be stratified based on the total number of test cases in the considered projects. In this way, we will analyze a sample that keeps the same proportion of test cases of the original population, \ie a larger project will account for more tests than a smaller one. 

As a first step, both inspectors will independently analyze a subset of 1,255 test methods (equal to 10\% of the total)---a third inspector (\ie the third author of the paper) will be in charge of making the final decision about the disagreements. Then, the results of this first validation will be compared through Cohen’s $\kappa$ statistic~\cite{cohen1960coefficient}, which measures the \emph{inter-rater agreement} of the inspection task. Through this measure, the inspectors will understand whether their manual analyses converge toward a common procedure, leading to an objective assessment. This step will be repeated, using different subsets, until a strong agreement is achieved~\cite{mchugh2012interrater}. As a final step of the validation process, the unclassified instances will be equally split between the two inspectors. This process will allow us to develop and deliver the largest manually-validated collection of test smells over 12,550 test cases. We will release the dataset as a publicly available source so that other researchers might exploit it to build on top of our findings.

\revised{While the formal process described above is supposed to mitigate possible bias when labeling the smelliness of test code, this may still contain subjective test smell instances. For this reason, we will also plan to involve experienced developers to validate the test smells affecting the considered tests. Since it seems unreasonable to ask for external validation of the entire set of 12,550 test cases (it would be excessively costly in terms of time and effort required by external developers), we will proceed as follows. We will randomly select a subset of 628 test cases (5\% of the test cases that will be validated) and involve 200 external developers through the \textsc{Prolific} platform,\footnote{\textsc{Prolific} website: \url{https://www.prolific.co/}.} a research instrument to select research participants. Through the appropriate filters, we will involve experienced developers. The developers will be provided with a definition of the test smells subject of the study and will be asked to assess the smelliness of four test cases. Note that, having 628 tests and 200 developers, we will be able to perform cross-checking, \ie several developers will assess a subset of 172 test cases to verify the consistency among the evaluations provided. In doing so, we will also collect background information about the participants. Should the external validation results be in line with or close enough to the internal one, we will consider the dataset construction completed. Otherwise, we will extend the external validation to the entire set of test cases by involving up to 600 developers recruited through \textsc{Prolific}.}

\section{Machine Learning-based Test Smell Detection}
\label{sec:approach}
This section illustrates the envisioned machine learning-based approach for test smell detection.

\begin{table*}
	\centering
	\caption{Test smells included in our study, their definition and the independent variables for each smell under investigation.}
	\label{tab:test-metrics}
		
		\begin{tabular}{c|l|c|p{5cm}}
			\hline
		    \rowcolor{gray20} \textbf{Test Smell} & \textbf{Definition} & \textbf{Metric} & \textbf{Description} \\\hline
			
			 & & NMC & Number of Method Calls \\
			\cline{3-4}
			 \multirow{2}{*}{Eager Test} & \multirow{2}{*}{A test method that invokes many methods of the object being tested.} & \multirow{2}{*}{PTMI} & Number of Production Types Method Invocations \\
			\cline{3-4}
			 & & \multirow{2}{*}{PET} & Probability of a Method to be affected by Eager Test based on its textual content \\
			\hline
			\rowcolor{gray10} \multirow{1}{*}{Mystery Guest} & \multirow{1}{*}{A test that uses external resources (\eg databases or files).}& NRF & Number of References to Files  \\
			\cline{3-4}
			\rowcolor{gray10} & & NRDB & Number of References to Database \\
			\hline
    		\multirow{3}{*}{Resource Optimism} & \multirow{3}{*}{A test that uses external resources without checking the state of these.} & \multirow{2}{*}{ERNC} & External Resource state (not files) Not Checked \\
            \cline{3-4}
            & & FRNC & File Resource state Not Checked \\
            \hline
		    \rowcolor{gray10} \multirow{4}{*}{Test Redundancy} & \multirow{4}{*}{A test that could be removed without impacting the test suite.} & \multirow{3}{*}{PR} & Pair Redundancy is the ratio between the items covered by a test and those covered by another one \\
		    \cline{3-4} 
		    \rowcolor{gray10} & & \multirow{3}{*}{SR} & Suite Redundancy is the ratio between the items covered by a test compared and those covered by all others tests in the test suite \\
		    \cline{3-4} 
			\hline
		\end{tabular}
\end{table*}


\subsection{Dependent Variable}
Our goal consists of automatically detecting the presence of test smells in test code components. Therefore, as a dependent variable, we will rely on a binary value indicating the presence/absence of a specific test smell type. We will consider as a dependent variable the outcome of the validation process discussed in \Cref{subsec:validation}.

\subsection{Independent Variables}
\label{sec:independent}
To collect a set of reliable predictors for each test smell under consideration, we will use the metrics from heuristic approaches already available in the literature.
Specifically, while performing the process described in \Cref{subsec:smell_selection} for the inclusion of test smells, we collected all the metrics defined and used by the available detection approaches. \Cref{tab:test-metrics} reports the list of metrics used for classifying each test smell with their description. Our online appendix~\cite{appendix} also includes references to all the tools relying on the same metrics. However, we are aware that these metrics might not represent a complete set of features characterizing test smells, \ie there might be additional metrics not considered by previous work that could contribute to identifying the four test smells. We will exploit the experience gained while manually identifying test smells to search for additional patterns and metrics that may be used as independent variables. We will reserve the right to define new metrics that complement the existing ones. 

\subsection{Selecting Machine Learning Algorithms}
Our work proposes the first machine learning-based test smell detector; therefore, the most suitable classifier is still unknown.
We will experiment with a set of classifiers belonging to different families that have been widely used in problems related to software maintenance and evolution~\cite{catolino2018enhancing,catolino2019extensive,catolino2019cross,di2017developer,pecorelli2019comparing,pecorelli2019role}.
The goal is to (i) understand which machine learning algorithm is the best for test smell detection and (ii) increase the generalizability of the results.
In details, we will evaluate \emph{Decision Tree}~\cite{freund1999alternating}, \emph{Naive Bayes}~\cite{ro1973pattern}, \emph{Multilayer Perceptron}~\cite{taud2018multilayer}, and \emph{Support Vector Machine}~\cite{noble2006support}, as basic classifier. We will also consider two ensemble techniques, such as \emph{Ada Boost}~\cite{schapire2013explaining} and \emph{Random Forest}~\cite{breiman2001random}.

\subsection{Configuration and Training}
\revised{When training the selected machine learners, we will experiment with multiple under- and over-sampling techniques to balance our data. We will compare them as further reported in \Cref{sec:rq2design}.}
As for the under-sampling, we will consider the use of \textsc{NearMiss 1}, \textsc{NearMiss 2}, and \textsc{NearMiss 3} algorithms~\cite{yen2006under}. 
Finally, we will experiment with a \textsc{Random Undersampling} approach that randomly explores the distribution of majority instances and under-samples them.
As for the over-sampling, we will experiment with \emph{Synthetic Minority Over-sampling Technique}, a.k.a SMOTE~\cite{chawla2002smote}, and advanced versions of this algorithm, \ie \emph{Adaptive Synthetic Sampling Approach}, a.k.a ADASYN~\cite{he2008adasyn} and the \textsc{Borderline}-SMOTE~\cite{han2005borderline}. 
We will also experiment with a \textsc{Random Oversampling} approach that randomly explores the distribution of the minority class and over-samples them.

Finally, concerning the classifiers configuration, we will experiment with the hyper-parameters of the classifiers using the \textsc{Random Search} strategy~\cite{bergstra2012random}: this search-based algorithm randomly samples the hyper-parameters space to find the best combination of hyper-parameters maximizing a scoring metric (\ie 
the Matthews Correlation Coefficient).
We plan to develop the entire pipeline with the \textsc{Scikit-Learn} library~\cite{scikit-learn} in \textsc{Python}. 

\subsection{\revised{Validation of the Approach}}
\revised{To assess the performance of our models, we will perform within- and cross-project validation. These validations aim to quantify the performance of the models in two different scenarios. We are interested to understand (i) how accurate can the performance be when a test smell detection model is trained using data of the same project where it should be applied and (ii) how accurate is the model when trained using external data to the project where it should be applied.}

\revised{For the within-project validation, we will perform a stratified 10-fold cross-validation~\cite{stone1974cross}---we will apply it to individual projects. This strategy randomly partitions the data into ten folds of equal size, allowing us to maintain the correct proportion in every split between smelly and non-smelly instances. It iteratively selects a single fold as a test set, while the other nine are used as a training set.}

\revised{For the cross-project validation, we will adopt the \emph{Leave-One-Out Cross-Validation} strategy~\cite{refaeilzadeh2009cross}, a special case of $K$-fold cross-validation with $K$ equal to $N$, the number of projects in the set. We will train models using the test cases of $N-1$ projects and use the test cases of the remaining project as the test set. The process will be repeated $N$ times to ensure that each project will occur in the test set once.}

\section{Execution Plan}
\label{sec:methodology}
\subsection{\textbf{RQ$_1$} - In Search of Suitable Metrics for Machine Learning-Based Test Smell Detection}
Finding a set of metrics to characterize the four considered test smells represents a first challenge to face~\cite{nicodemus2009predictor}.
As explained in \Cref{subsec:smell_selection}, we will start focusing on the metrics that have been used by previous researchers when detecting test smells.
In other words, we will investigate whether a machine learning solution is suitable to combine structural and textual metrics that were considered in isolation by previous work. 
\Cref{tab:test-metrics} lists and describes each considered test smell.
These metrics capture the smelliness of tests under different perspectives, taking into account the size of fixtures and test suites, cohesion and coupling aspects of tests, and conceptual relationships between the methods composing test suites.

\revised{We will quantify the predictive power of each metric by computing their \emph{information gain}~\cite{quinlan1986induction}. This step will be used as a \emph{probing} method, \ie it will estimate the contribution provided by each metric other than acting as a feature selection instrument: we will indeed use as predictors the metrics having an information gain higher than zero, i.e., we will discard the metrics that do not provide any expected beneficial effect on the performance.}
More specifically, the output of the information gain algorithm consists of a ranked list where the features of the model are placed in a descending manner, meaning that those contributing the most are placed at the top.
We will employ the \emph{Gain Ratio Feature Evaluation} algorithm \cite{quinlan1986induction} available in the \textsc{Scikit-Learn} library~\cite{kramer2016scikit}. 


\subsection{\textbf{RQ$_2$} - Assessing the Performance of Our Machine Learning-Based Test Smell Detector}
\label{sec:rq2design}
\revised{When assessing the performance of our models, we will proceed with a stepwise analysis of the various components included in the experimentation. We will perform an \emph{ablation} study to analyze the contribution of each configuration and training step to the overall models' performance. We will experiment with multiple combinations, \eg we will experiment how the performance varies when including (and not) the feature selection step, the data balancing, and the hyper-parameter optimization, other than considering the performance variations given by the adoption of different validation procedures. In this way, we will also be able to assess the best possible pipeline for the problem of test smell detection.}
		
\revised{To evaluate the performance of the various combinations experimented and address \textbf{RQ$_2$}, we will compute a number of state-of-the-art metrics such as \emph{precision}, \emph{recall}, \emph{F-Measure}~\cite{baeza2011modern}, \emph{Matthews Correlation Coefficient} ($MCC$) \cite{baldi2000assessing}, and the \emph{Area Under the Curve - Precision-Recall (AUC-PR)}.}

To support the results achieved, we will statistically verify the validity of the findings.
We will use the Wilcoxon test~\cite{wilcoxon1945individual}, with 0.05 as a significance value, computed on the distributions of MCC values of machine learning-based and heuristic-based techniques over the different projects and the different test smell types.
We will also rely on Cliff's Delta (or $d$), a non-parametric effect size measure~\cite{cliff1993dominance} to assess the magnitude of the measured differences. 

\subsection{\textbf{RQ$_3$} - Comparing Machine Learning- and Heuristic-Based Techniques for Test Smell Detection}

\revised{To complement the analysis of the performance of our machine learning-based approach, we plan to conduct a benchmark study to compare our approach with state-of-the-art techniques based on heuristics. On the one hand, we will assess the real usefulness of our approach: should our model be less performing than the baselines, its practical use would be limited. On the other hand, we will measure the extent to which our technique overcomes existing approaches, thus understanding the strengths and weaknesses of our approach compared to existing detectors. 
We will compare our approach against three heuristic-based baselines:}

\begin{description}[leftmargin=0.3cm]
	

    \item[\textsc{tsDetect}~\cite{peruma2020tsdetect}.] \revised{We select this tool as it represents the current state of the art in test smell detection \cite{aljedaani2021test} and, at the same time, it is able to detect the highest number of test smell types. Out of the four test smells included in our study, \textsc{tsDetect} can identify three of them, \ie \emph{Eager Test}, \emph{Mystery Guest}, and \emph{Resource Optimism}. In particular, the first is detected by computing the number of the multiple calls made by a test method to multiple production methods. 
    The second is identified by analyzing whether a test method contains instances of files and database classes. Finally, the third is identified by looking at whether a test method utilizes a \texttt{File} instance without calling the method \texttt{exists()}, \texttt{isFile()}, or \texttt{notExist()}.}
    
    \smallskip
    \item[\textsc{TeReDetect}~\cite{koochakzadeh2010tester}.] \revised{We select this tool as it is the only one available to detect \emph{Test Redundancy} smell instances. The tool detects the smell by computing code coverage and analyzing whether two tests cover similar paths.} 
	
	\smallskip
	\item[\textsc{Darts}~\cite{lambiase2020just}.] \revised{The model built for \emph{Eager Test} relies on an information retrieval metric (\ie PET). For this reason, we believe it might be worth to compare the model against an information retrieval-based heuristic technique, which is the one implemented within \textsc{Darts}~\cite{lambiase2020just}. The tool relies on the detection rule proposed by Palomba \etal~\cite{palomba2018automatic}. As such, it detects \emph{Eager Test} instances through a two-step process: first, the test method calls are replaced with the actual production code methods called by the test method; then, the conceptual cohesion metric is computed, taking into account the constituent methods and, whether this metric exceeds 0.5 the smell is detected.}

\end{description}

\revised{To enable a fair comparison, we will run the heuristic approaches against the same systems considered in \textbf{RQ$_2$}. None of these heuristic tools require configuration, \ie they can be run against the source code without the need of specifying any parameter: this ensures the execution of their original implementations, hence avoiding possible bias due to wrong configuration of the tools. We will employ the same evaluation metrics used to assess the machine learning models, \ie \emph{precision}, \emph{recall}, \emph{F-Measure}, \emph{MCC}, and \emph{AUC-PR}. Similarly to \textbf{RQ$_2$}, we will also statistically verify the validity of the findings between our and baseline techniques by using the Wilcoxon Test~\cite{wilcoxon1945individual} and the Cliff's Delta~\cite{cliff1993dominance}.}

\subsection{\revised{\textbf{RQ$_4$} - In-Vivo Evaluation of the Machine Learning-Based Test Smell Detector}}

\revised{After assessing the performance of the machine learning-based approach through lab experimentation and comparison with the state of the art, we will then consider the definition of a new empirical analysis aiming at verifying the capabilities of the model in the wild. More specifically, we plan to experiment the machine learning models against the test code of open-source systems not included in the dataset. We will (i) collect the 50 most popular GitHub projects - according to their number of stars; (ii) apply the models against the test code of those projects to detect test smells; (iii) open new issues on their issue tracker, one for each test smell detected; (iv) assess how developers react to these pieces of information. Should these predictions be used by developers to discuss about the quality of their tests or to refactor them, this would imply that our model is effective in practice. We will compute metrics such as the number of comments per issue, the number of refactoring actions performed by developers, and the number of closed issues. In addition, we will also report on the developer's opinions expressed in the comments. With this \emph{in-vivo} evaluation we aim at (i) measuring the extent to which the predictions of the models are actually usable and useful for developers to identify and remove potential issues in test code; and (ii) reducing possible threats to generalizability. In any case, it is worth remarking that such an evaluation would be worth to be conducted \emph{if and only if} the performance of the models would be high enough; otherwise, the in-vivo validation would become an unnecessary waste of the time for developers, who would be dealing with noise in their issue trackers. Hence, we will decide on whether to pursue the experiment based on the F-Measure achieved in \textbf{RQ$_2$}: should this be higher than 70\%, we will proceed with the evaluation.}

\subsection{Publication of Generated Data}

All the data generated from our study will be publicly available in an online repository~\cite{appendix}. We also plan to release the scripts, other than the data collected and used for the statistical analysis that we will present in the paper.

\section{Limitations}
\label{sec:ttv}
\revised{A first possible limitation will concern the dataset exploited in our study. Starting from a publicly available dataset containing 51,549 test cases, we will manually identify test smells on a stratified sample of 12,550 tests. Such a manual inspection will represent the main threat to the validity of our conclusions. To mitigate this threat, we plan to involve more inspectors who will follow a systematic approach. First, they will independently analyze subsets of test cases and repeatedly compute inter-rater agreement measures to verify to what extent the manual inspection will converge toward a common procedure. Only after this step, they will analyze the remaining unclassified test cases. Whenever needed, a third inspector will also help solve disagreements. In addition, we also plan to involve external, experienced developers within an additional manual validation of a sample of test cases. Should it not align with our internal validation, we will proceed with a full external validation based on the developers' assessment.}

\revised{In \textbf{RQ$_3$}, we selected alternative state-of-the-art heuristic approaches which have been employed by the research community, showing significant results (\eg like in the case of \textsc{TsDetect} \cite{peruma2019distribution,peruma2020exploratory} and \textsc{Darts} \cite{lambiase2020just}), or the only available tool for the detection of \emph{Test Redundancy} (\ie \textsc{TeReDetect} \cite{koochakzadeh2010tester}). In this respect, it is worth remarking that none of them require configuration; therefore, we can rely on their original implementations.}
\revised{Other limitations may involve the implementation of our machine learning-based detector. From a methodological standpoint, we will conduct \emph{probing} and \emph{ablation} studies to identify the most relevant independent variables and the most appropriate pipeline to devise the models. As such, possible threats to the creation of the models will be mitigated by analyzing multiple aspects that might influence the results (\eg which features to consider and how to train the classifier). We believe that the procedures we will follow are precise enough to ensure the validity of the study. From a technical perspective, we will rely on the \textsc{Scikit-Learn} library~\cite{scikit-learn}, thus leveraging all its structures and algorithms. While we cannot exclude possible implementation errors, the development community around \textsc{Scikit-Learn} usually tests source code appropriately.} 

Concerning the relationship between treatment and outcome, we will exploit a set of widely-used metrics to evaluate the experimented techniques (\ie precision, recall, F-measure, MCC, AUC-PR), and we will provide qualitative examples to show the differences between the compared approaches. \revised{When assessing the contribution of the features to use in our approach, we will rely on the \emph{Gain Ratio Feature Evaluation} algorithm~\cite{quinlan1986induction}, which the research community has widely used for the same purpose~\cite{catolino2018enhancing,catolino2019cross,palomba2017toward}.} Finally, we will use appropriate statistical tests, \ie the Wilcoxon Test and the Cliff's Delta, which will allow us to support our findings.

\revised{In terms of generalizability, we will validate the devised models in both within- and cross-project scenarios to assess the capabilities of the models at large. Especially with the cross-project validation, we expect to provide insights into the generalizability of the results. At the same time, should the performance of the devised models be sufficiently high, we will experiment with them with the test code of open-source systems not included in the dataset. Such an \emph{in-vivo} evaluation is in line with the \emph{lab-to-field generalization} strategy proposed by Wieringa and Daneva \cite{wieringa2015six}. In this generalization strategy, a technology (\eg the prediction models) is first experimented with in-house (\eg through within- and cross-project validation) and then assessed in the field where it is supposed to be used (\eg any arbitrary open-source systems). Nonetheless, we still recognize the presence of additional threats to generalizability. These are mainly connected to (i) our focus on \textsc{Java} projects, which is due to the availability of tools working for this programming language \cite{aljedaani2021test} and (ii) the large, but still limited, number of projects that we will consider. In this respect, we can claim for the \emph{generalizing by similarity} principle described by Ghaisas \etal \cite{ghaisas2013generalizing}: it is likely that similar results might be obtained in projects having similar characteristics to those analyzed in our work.}

\section{Conclusion}
\label{sec:conclusion}
The ultimate goal of our research is to define a machine learning-based test smell detection and compare its performance with those of heuristic baselines. We will start working toward this goal by creating the largest manually-validated dataset of test smells: we will report information about the presence of four smell types over 12,550 test cases.
As part of our future work, we plan to assess our technique following the methodology described in \Cref{sec:methodology}. 

\section*{Acknowledgement}
Fabio gratefully acknowledges the support of the Swiss National Science Foundation through the SNF Projects No. PZ00P2\_186090. This work has been partially supported by the EMELIOT national research project, funded by the MUR under the PRIN 2020 program (Contract 2020W3A5FY).

\balance
\scriptsize
\bibliographystyle{abbrv}
\bibliography{main}  

\end{document}